# Local and regular plasma oscillations in bulk donor type semiconductors

Yuri Kornyushin

Maître Jean Brunschvig Research Unit, Chalet Shalva, Randogne, CH-3975

**Abstract**

Restoring force acts on the electronic cloud of the outer electrons of a neutral or charged impurity atom when it is shifted relative to the inner charged core. Because of this the dipole oscillation arises, which influences considerably the dispersion law of the plasma oscillation in bulk donor semiconductors. Assuming that only one transition of the outer electron from the ground state to the first excited state is essential, the dispersion law is calculated. It is shown that calculated dispersion law consists of two separate branches, one of them originates from the regular plasma oscillation of the free electrons of a conductivity band, and the other one stems from the local oscillation of the outer electrons bounded to the impurity atoms.

## 1. Introduction

The mass of an impurity atom in a crystal is different from that of the host atom. The interaction of an impurity atom with the host atoms is also different from that of the host atoms. This influences the frequency of the oscillation of an impurity atom. When the frequency of the oscillation of an impurity atom is found outside the phonon spectrum of a crystal, the oscillation is called a local oscillation [1]. When the frequency of an impurity atom is found inside the phonon spectrum of a crystal, this oscillation is called a quasilocal oscillation [1]. Local and quasilocal oscillations of phonon modes in crystals, containing light and heavy impurity atoms are well known and have been studied in details [1]. Local oscillation manifests itself also in plasmon modes of semiconductor superlattices. Bulk and surface local plasmon modes of the semiconductor superlattices were studied in [2,3,4].

Here let us consider local oscillation of the electronic cloud around neutral donor impurity atoms in bulk semiconductors and its interaction with regular plasma oscillation of the current carriers [5]. Regular plasma oscillation of the current carriers is accompanied by the oscillating electric field, which influences other oscillations in a sample. This interaction changes significantly the dispersion law of the collective electron oscillations. Interaction of the local oscillation of the electronic cloud around neutral impurity atoms with the regular plasma one is the simplest example of the interaction between different modes of oscillations. It is possible when the frequency of the local oscillation is slightly higher than the Langmuir frequency.

## 2. Hydrogen atom model for donor impurity atom

The donor impurity atom has usually one more outer electron comparative to the host atoms of a semiconductor (e. g., the silicon atoms). Therefore point positively charged core and one electron could model it. When the impurity atom is ionized, the electron moves to the conductivity band. When this extra electron is bound to the donor impurity atom, the hydrogen atom in a dielectric media with dielectric constant $\varepsilon_0$ can model it. This approximation is reasonable when the size of the electronic cloud $2R = \varepsilon_0 \hbar^2/me^2$ [5], is large enough, and it is so for semiconductors with the dielectric constant $\varepsilon_0$ larger than 5 (quite typical value of $\varepsilon_0$ is about 10). In this case the size of the atom $2R$ comprises several lattice parameters. To consider the approximation discussed reasonable one should also assume that the screening (Debye for the Boltzmann statistics or Thomas-Fermi for the degenerated current carriers) radius is much larger than $R$. In this case the current carries do not influence significantly formation of a "hydrogen atom". In this model the wave function of the electron in the ground state is [6]

$$\psi(r) = [\exp(-r/2R)]/2(2\pi)^{1/2}R^{3/2}, \tag{1}$$

where $r$ is the radius-vector.

The energy of the ground state is considerably smaller ($\varepsilon_0^2$ times) for the hydrogen atom in a dielectric media and it is described by the following equation [6]:

$$E_0 = -(me^4/2\varepsilon_0^2\hbar^2), \tag{2}$$

where $e$ is the electron (negative) charge, $m$ is the effective electron mass, and $\hbar$ is the Planck constant divided by $2\pi$

## 3. Interaction of bulk and local plasma oscillations

In quantum mechanics the energy levels of the electron in a hydrogen atom are [6]

$$E_\nu = -(me^4/2\varepsilon_0^2\nu^2\hbar^2), \ \nu = 1, 2, 3, \ldots. \tag{3}$$

The transitions between these levels cause the polarization of the impurity atom when external electric field is applied [6]. External alternative electric field excites regular plasma oscillation of the current carriers in a semiconductor. In a semiconductor regular plasma oscillation of the current carriers is accompanied by the oscillation of the electric field of the same frequency inside a sample. This field acts on the impurity atoms along with the external one. Thus regular plasma oscillation of the current carriers interacts with the polarization of impurity atoms. This interaction leads to a modification of the dispersion law of the oscillations. Let us assume further that only one transition, transition from the ground level to the first excited one, should be taken into account, and that the other transitions could be neglected.

The dielectric permeability of a semiconductor sample as a function of the angular frequency $\omega$ and the wave vector $k$ is usually regarded to be as follows [7]:

$$\varepsilon(\omega,k) = 1 - (\omega_{pi}/\omega)^2 + 4\pi\chi, \tag{4}$$

where $\omega_{pi}^2 = (4\pi e^2 n/m) + a_0 k^2$ is the initial plasma frequency of the electronic gas of the conductive electrons of a density $n$ without taking into account the specific polarizability of the substance of a sample $\chi$ without free conductive electrons. When the oscillation frequency of the electrons is much higher than the collision frequency then, as was shown in [8], the oscillations of the electrons are essentially one-dimensional and adiabatic. In this case the factor $a_0 = 3(k_B T/m)\varepsilon_0$ for the case of the Boltzmann statistics, or $a_0 = 1.2(E_F/m)\varepsilon_0$ for a degenerated electronic gas ($k_B$ is the Boltzmann constant, $T$ is the absolute temperature, and $E_F$ is the Fermi energy). When the oscillation frequency of the electrons is much lower than the collision frequency, then $a_0 = (k_B T/m)\varepsilon_0$ for the case of the Boltzmann statistics, or $a_0 = (2E_F/3m)\varepsilon_0$ for a degenerated electronic gas [8].

Taking into account that $1 + 4\pi\chi = \varepsilon_0$, and that the polarizability of the bound electrons of the donor impurity atoms which are not ionized [7],

$$\alpha_i = f n_d e^2/m(\omega_0^2 - \omega^2) \tag{5}$$

($f$ is so called oscillatory force, $n_d$ is the number of the neutral, not ionized, donor atoms per unit volume of a semiconductor, and $\omega_0 = 3me^4/8\varepsilon_0^2\hbar^3$ is the frequency of the transition from the ground state to the first excited one, due to which the polarizability arises), one can come to the following expression for the dielectric permeability as a function of the angular frequency and the wave vector:

$$\varepsilon(\omega,k) = \varepsilon_0\{1 - (\omega_p/\omega)^2 + [\omega_d^2/(\omega_0^2 - \omega^2)]\}, \tag{6}$$

where $\omega_p^2 = \omega_{p0}^2 + ak^2$ is the square of the regular plasma frequency of the free conductive electrons in a semiconductor, $\omega_{p0}^2 = (4\pi e^2 n/\varepsilon_0 m)$ is the square of the Langmuir frequency, $a = a_0/\varepsilon_0$, and $\omega_d^2 = 4\pi f n_d e^2/m$.

As was mentioned above, in Eq. (6) one transition from the ground state to the first excited state only was taken into account.

The dispersion law, that is the dependence of the angular frequency on the wave vector, is determined in quantum mechanics by a well known condition $\varepsilon(\omega,k) = 0$ [7]. From this and Eq. (6) follows that there are two branches of the oscillations:

$$\omega_{1,2}^2 = 0.5(\omega_{p0}^2 + \omega_0^2 + \omega_d^2 + ak^2) \pm 0.5[(\omega_{p0}^2 - \omega_0^2 + ak^2)^2 + \omega_d^2(2\omega_{p0}^2 + 2\omega_0^2 + \omega_d^2 + 2ak^2)]^{1/2}. \tag{7}$$

At $\omega_d^2 = 0$ one has $\omega_1 = \omega_p = [(4\pi e^2 n/\varepsilon_0 m) + ak^2]^{1/2}$ and $\omega_2 = \omega_0$. This means that at low concentration of the neutral donor impurity atoms one has two separate modes of the oscillations considered: a regular plasma oscillation of the free electrons in a conductivity band and a local plasma oscillation of the outer electrons bounded to the donor impurity atoms. When $\omega_d^2 = 0$ and $\omega_0$ is slightly larger than $\omega_{p0}$ the two branches of the oscillations cross at some $k$. But when $\omega_d^2$ is not zero, one can see from Eq. (7) that the two branches never cross. One of them goes below the other one. This means that in this case the dispersion law changes considerably. Typical dispersion curves are shown in Fig. 1. They were calculated using Eq. (7) for $\omega_0 = 1.1\omega_{p0}$ and $\omega_d^2 = 0.1\omega_{p0}^2$.

A similar problem, involving interaction between bulk plasmons and bulk longitudinal optic phonons, was studied theoretically by A. Mooradian et al [9,10] in the same approach as in the present paper (optic phonons and free electrons make additive contributions to the total dielectric response function, optic phonon contribution has a pole at the transverse phonon frequency). The effect of anticrossing between the two interacting modes similar to that depicted in Fig. 1 was predicted in those papers. Longitudinal optic phonon and plasmon coupling in bulk doped GaAS was also studied experimentally by K. Kuriyama [11]. So the calculations presented in this paper are not essentially new, but the physics is different. The dependence of obtained results on the basic parameters of semiconductor (effective mass, dielectric constant, etc.) is also different.

The condition of the existence of the regarded anticrossing is that the frequency of the local plasma oscillation $\omega_0$ should be slightly larger than that of the Langmuir oscillation $\omega_{p0}$. From this follows [taking into account that $\omega_0 = 3me^4/8\varepsilon_0^2\hbar^3$ and $\omega_{p0} = (4\pi e^2 n/\varepsilon_0 m)^{1/2}$] that when $n$ is slightly larger than

$$n_0 = 9m^3 e^6/256\pi\varepsilon_0^3\hbar^6, \qquad (8)$$

the regarded anticrossing takes place. For example when $m = 0.453 m_e$ and $\varepsilon_0 = 11.7$ Eq. (8) yields $n_0 = 4.34\times 10^{18}$ cm$^{-3}$, the Debye screening radius, $g^{-1} = (\varepsilon_0 kT/4\pi e^2 n_0)^{1/2} = 9.42\times 10^{-5}$ cm. This value of $g^{-1}$ is orders of magnitude larger than the radius of the hydrogen atom in a dielectric medium. For $m = 0.453 m_e$, $\varepsilon_0 = 11.7$, and $n = n_0 = 4.34\times 10^{18}$ cm$^{-3}$ the Langmuir frequency $\omega_{p0} = (4\pi e^2 n/\varepsilon_0 m)^{1/2} = 5.1\times 10^{11}$ 1/s.

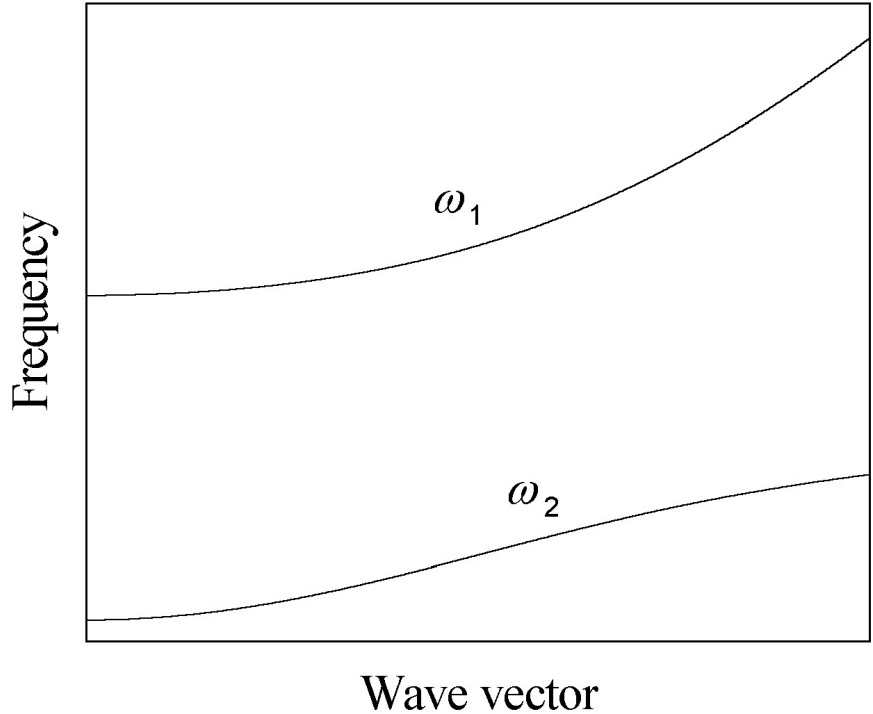

*Fig. 1.* Two branches of a typical dispersion curve, calculated using Eq. (7) for $\omega_0 = 1.1\omega_{p0}$ and $\omega_d^2 = 0.1\omega_{p0}^2$.

At high enough concentration of neutral donor impurities the impurity levels of the oscillation form so-called impurity band [1]. When there are no free electrons in the conductivity band and there is no regular plasma oscillation, Eq. (7) yields that the local mode only is present.

The acceptor impurity atom may have also one or two outer electrons, which contribute to the formation of a local mode of the plasma oscillation in the same manner as was considered above. Only the factors in equations have their specific values.

So, it is worthwhile to mention that the dipole oscillation of the electronic cloud of the impurity atoms in bulk semiconductors interacts with the regular plasma oscillation of the conductive electrons. This leads to a significant modification of the dispersion laws of the plasma oscillations. It was shown that as a result of this interaction the two branches of the plasma oscillation appear in the bulk extrinsic semiconductor.

## 4. Conclusions

From the results of the calculations performed here it is possible to draw following conclusions:

In a bulk semiconductor of a donor type, when the concentration of conductive electrons is high enough, the frequency of the oscillation of the outer electrons of neutral donor impurities (local plasma oscillation) may exceed slightly the Langmuir frequency. This makes anticrossing of the regular branch of the plasma oscillation with the local oscillation branch possible. Regarded anticrossing leads to the considerable changes of the dispersion law of the plasma oscillations.